\RequirePackage[colorlinks,allcolors=blue,pdfborder={0 0 0}]{hyperref}
\documentclass[draft]{agujournal2019}
\usepackage{url}
\usepackage{lineno}
\usepackage{soul}
\usepackage{amsmath}
\usepackage{siunitx}

\usepackage[final]{pdfpages}

\drafttrue

\journalname{}

\begin{document}

\title{Centennial-scale variability of the Atlantic Meridional Circulation in CMIP6 models shaped by Arctic–North Atlantic interactions and sea ice biases}

%
%
\authors{Oliver Mehling\affil{1}, Katinka Bellomo\affil{1,2}, and Jost von Hardenberg\affil{1,2}}

\affiliation{1}{Department of Environment, Land and Infrastructure Engineering, Politecnico di Torino, Turin, Italy}
\affiliation{2}{National Research Council of Italy, Institute of Atmospheric Sciences and Climate (CNR-ISAC), Turin, Italy}

\correspondingauthor{Oliver Mehling}{oliver.mehling $\langle$at$\rangle$ polito.it}

\begin{keypoints}
\item We present the first systematic multi-model comparison of internal centennial-scale AMOC variability in state-of-the-art climate models
\item A robust mechanism of Arctic--North Atlantic freshwater exchange is identified only in models that use the NEMO ocean component
\item Sea ice cover biases in convective regions of the North Atlantic amplify AMOC variability and could provide an observational constraint
\end{keypoints}

%
%

\begin{abstract}
Climate variability on centennial timescales has often been linked to internal variability of the Atlantic Meridional Overturning Circulation (AMOC). However, due to the scarceness of suitable paleoclimate proxies and long climate model simulations, large uncertainties remain on the magnitude and physical mechanisms driving centennial-scale AMOC variability. For these reasons, we perform, for the first time, a systematic multi-model comparison of centennial-scale AMOC variability in pre-industrial control simulations of state-of-the-art global climate models. Six out of nine models in this study exhibit a statistically significant mode of centennial-scale AMOC variability. Our results show that freshwater exchanges between the Arctic Ocean and the North Atlantic provide a plausible driving mechanism in a subset of models, and that AMOC variability can be amplified by ocean–sea ice feedbacks in the Labrador Sea. The amplifying mechanism is linked to sea ice cover biases, which could provide an observational constraint for centennial-scale AMOC variability.
\end{abstract}

\section*{Plain Language Summary}
\noindent Changes in ocean circulation are often proposed as drivers of natural variations of the Earth's climate on timescales of centuries. However, it is unclear how strong these natural variations of the circulation strength, called internal variability, are in the real world, because reconstructions from the past climate are sparse and climate models are expensive to run for these long timescales. Here, we compare how the latest generation of climate models simulate internal variability of the Atlantic Meridional Overturning Circulation (AMOC) – the ocean circulation that is often thought to be responsible for Europe’s comparatively mild climate – on timescales of 100 to 250 years. We find that many models have stronger variability on these timescales than what would be expected simply from random noise. In several models, AMOC variability appears to be driven by the release of fresh water from the Arctic Ocean and amplified by intermittent sea ice cover in the North Atlantic. However, this amplification only occurs if a model simulates a too extensive sea ice cover in winter. This mechanism shows that sea ice cover -- which is easily observable -- could be used to constrain variability of the AMOC on timescales longer than the observational record.

%
%

\section{Introduction}
\noindent Past and future climate change is determined by both external forcing (such as increasing anthropogenic CO$_2$ emissions or volcanic eruptions) and internal variability that arises from chaotic interactions between the different components of the climate system. Hence, assessing the magnitude of and mechanisms responsible for internal variability is crucial for regional climate projections \cite<e.g.,>{Lehner2020}, detection and attribution \cite{Eyring2021}, and the interpretation of the paleoclimatic record \cite{VonDerHeydt2021}. While variability on interannual to decadal timescales can often be studied by combining large ensembles of climate models and the instrumental record, for longer timescales the uncertainty is much larger \cite{Laepple2023} due to limited climate reconstructions and the computational cost of long climate model integrations.

Here, we focus on modes of climate variability on centennial timescales (defined as a period of 100--250 years), which have often been linked to internal variability of the Atlantic Meridional Overturning Circulation (AMOC) \cite<e.g.,>{Knight2005,VonDerHeydt2021,Ellerhoff2022,Bakker2022}. Centennial-scale AMOC variability has been studied less extensively than the neighboring multidecadal timescales \cite{Buckley2016}, but might still imprint on the climate at human timescales \cite<e.g.,>{Bonnet2021a,Kelson2022}. Because the AMOC strength is challenging to reconstruct from available sea surface temperature proxies \cite{Moffa-Sanchez2019,Little2020,Bakker2022}, and circulation proxies often do not provide sufficient resolution \cite{Lippold2019}, here we focus on simulated centennial-scale AMOC variability in state-of-the-art climate models.

In single climate model studies, several different mechanisms for a centennial-scale mode of AMOC variability have been suggested. Proposed drivers include the propagation of salinity anomalies from the southern hemisphere \cite{Delworth2012,Martin2015}, subtropical precipitation anomalies \cite{Vellinga2004}, freshwater transport from the Arctic Ocean \cite{Jiang2021,Meccia2023,Mehling2023}, and internal ocean mixing feedbacks in the North Atlantic \cite{Li2022,Prange2023,Yang2024a}. This diversity demonstrates a need for systematic model intercomparison of centennial-scale AMOC variability and its mechanisms, which -- in contrast to shorter timescales \cite{Ba2014,Muir2015} -- has so far only been achieved with one very small (three-model) ensemble \cite{Menary2012}.

Here, we provide the first systematic intercomparison of centennial-scale AMOC variability in the latest generation of global climate models, making use of the unprecedented availability of long pre-industrial control (piControl) simulations in the Coupled Model Intercomparison Project Phase 6 \cite<CMIP6;>{Eyring2016}. We also compare the link between Atlantic overturning and freshwater exchanges with the Arctic Ocean, which has previously been proposed as a driving mechanism of centennial-scale AMOC variability in two of these CMIP6 models \cite{Jiang2021,Meccia2023}. Finally, we discuss inter-model diversity with a focus on sea ice biases in the pre-industrial mean state, which may help constrain simulated centennial-scale variability.

\section{Materials and Methods}
\subsection{CMIP6 model data}
\noindent To analyze internal variability, we use piControl simulations from CMIP6, in which the external forcing is held constant at 1850 levels \cite{Eyring2016}, hence the time evolution is governed by internal dynamics. We select the longest piControl simulation for each model if it spans at least 1000 years. This is to both sufficiently sample centennial-scale variability, and to separate internal variability from a residual model drift. For our analysis, we require that models provide at least the meridional overturning streamfunction (msftyz or msftmz), salinity (so), velocity (uo and vo), mixed layer depth (mlotst) and sea ice concentration (siconc) as output.

This yields a set of 9 models from 8 different modeling centers (Table S1), a small but diverse sample of the larger CMIP6 ensemble. All models analyzed here have a nominal ocean resolution of around \ang{1} and therefore parametrize mesoscale ocean eddies. However, in contrast to CMIP5, all models resolve two ocean gateways west of Greenland, allowing for a more consistent (and more realistic) representation of Arctic--North Atlantic linkages \cite{Zanowski2021}. Following \citeA{Jiang2021}, we detrend all time series quadratically to account for (potentially non-linear) model drift.

\subsection{Diagnostics}
\noindent We define AMOC strength for each latitude as the maximum of the Atlantic meridional overturning streamfunction over depth \cite{Buckley2016} below 500 m. Freshwater content in the Arctic Ocean is expressed in terms of the thickness of the water column above a reference salinity $S_\text{ref}$ \cite{Haine2015}:
\begin{equation}
    h_\text{fw}(x,y,t) = \int_{D(S_\text{ref})}^{0} \frac{S_\text{ref} - S(x,y,z,t)}{S_\text{ref}}\: \mathrm{d}z.
    \label{eq:fw-content}
\end{equation}
and freshwater transport into the Arctic through each strait is defined as
\begin{equation}
    \Phi_\text{fw} = \iint \mathbf{u}\, \frac{S_\text{ref}-S}{S_\text{ref}} \cdot \mathrm{d} \mathbf{A},
    \label{eq:fw-flux}
\end{equation}
where $\mathbf{u}$ is the velocity across a section of area $\mathrm{d} \mathbf{A}$ (pointing into the Arctic Ocean) \cite[and references therein]{Zanowski2021}. The integral is taken over the full ocean depth and horizontal extent of the strait. Sections are calculated on the native model grids, using the definitions of \citeA{Zanowski2021} where applicable.

Here, we choose the reference salinity as the volume-averaged Arctic Ocean mean salinity, delimited by the straits shown in Supplementary Fig. S1, for each model (Table S1). This approach has been taken in previous modeling studies \cite<e.g.,>{Cornish2020,Mehling2023} to account for different salinity biases of individual models. We tested that our results hold for the frequently used value of $S_\text{ref}=34.8$ and are therefore not sensitive to the exact choice of $S_\text{ref}$. Defining the fingerprint in Fig. \ref{fig:freshwater-correlation} through freshwater content instead of depth-averaged salinity anomalies \cite{Jiang2021,Meccia2023} yields a similar picture in the Arctic Ocean but avoids choosing an arbitrary reference depth as well as spuriously large anomalies in regions with shallow bathymetry.

Power spectra and coherency are computed using the multi-taper method \cite{Thomson1982,Percival2020}. To detect peaks in spectral power, we compare the spectra to the null hypothesis of a red noise spectrum generated by a first-order autoregressive (AR(1)) process \cite{Mann1996}. This method relies on smoothing the power spectrum before fitting an analytical AR(1) spectrum. Following the recommendations of \citeA{Mann1996}, we choose the smoothing bandwidth parameter as $\Delta f_\text{smooth} = 0.05\,\text{year}^{-1}$, which yields a good overall match between the fit and the smoothed spectra. Our results are not sensitive to reasonable variations of $\Delta f_\text{smooth}$. For lagged regressions, we test significance using the method of \citeA{Ebisuzaki1997}, controlling for multiple comparisons by using effective degrees of freedom \cite{Mudelsee2014} based on the autocorrelation of both timeseries.

\section{Results}
\subsection{Centennial-scale AMOC variability}

\begin{figure}[htbp]
    \centering
    \includegraphics[width=0.95\textwidth]{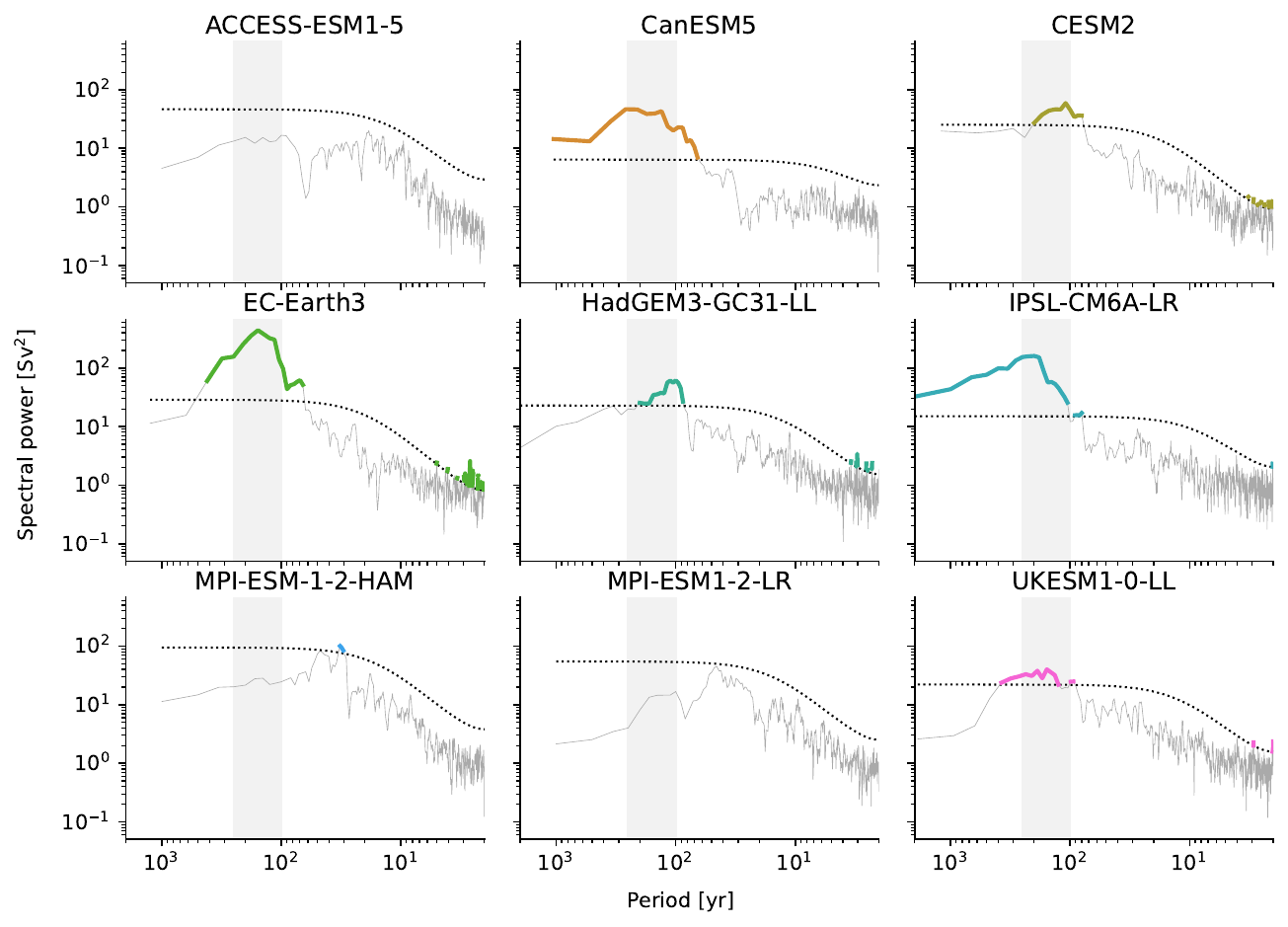}
    \caption{\textbf{Power of centennial-scale AMOC variability.} Multi-taper power spectra of detrended AMOC strength time series at \ang{40}N for CMIP6 models with at least 1000 years of piControl. Colored bands exceed the 99\% confidence level (dotted line) of the AR(1) fit. Gray shading indicates the centennial timescale of interest in this paper (period 100--250 years).}
    \label{fig:amoc-spectra}
\end{figure}

\noindent First, we use power spectral analysis to show that a significant mode of variability in AMOC strength can be identified in several long control simulations of CMIP6 models. Since most models exhibit the strongest centennial-scale AMOC variability in depth space at around \ang{40}N (Supplementary Fig. S2), we use the AMOC strength at \ang{40}N to characterize AMOC variability in the following. However, on centennial timescales, the AMOC at \ang{40}N is highly coherent (coherency $>0.92$ for all models in this study except ACCESS-ESM1-5) with the commonly used AMOC index at \ang{26.5}N (Supplementary Fig. S3), such that the results do not strongly depend on the exact choice of latitude for the AMOC index.

Fig. \ref{fig:amoc-spectra} shows the multi-taper power spectra of AMOC strength at \ang{40}N as a function of the period, with the timescales of interest (period 100--250 years) highlighted in gray. Compared to the null hypothesis of an AR(1) process, six of the nine analyzed models exhibit a significant mode of AMOC variability on centennial timescales at the 99\% confidence level. This includes all five models (IPSL-CM6A-LR, EC-Earth3, HadGEM3-GC31-LL, UKESM1-0-LL, CanESM5) that use NEMO as their ocean component (Table S1).

However, the amplitude and period of the peak spectral power of AMOC variability vary widely among models, with EC-Earth3 and IPSL-CM6A-LR showing stronger variability than the other models of the ensemble. Few models with a significant mode of variability show a clearly defined spectral peak at one timescale, but rather significant power across most of the 100--250 year range. Nevertheless, oscillations -- although perhaps not as regular as for EC-Earth3 and IPSL-CM6A-LR -- can be seen in the low-pass filtered time series in Supplementary Fig. S4 for all models with significant centennial-scale variability. Among these models, the standard deviation of the 70-year low-pass filtered time series ranges from \SI{0.5}{Sv} for UKESM1-0-LL to \SI{1.4}{Sv} for EC-Earth3. In the three remaining models, the low-pass filtered standard deviation is below \SI{0.5}{Sv}.

\subsection{Arctic--North Atlantic freshwater exchanges} \label{sec:fw-exchanges}
\noindent To gain a mechanistic understanding of the drivers of this centennial-scale AMOC variability across climate models, we focus on one mechanism, Arctic--North Atlantic freshwater exchanges. This mechanism has been proposed to drive the AMOC oscillations in the IPSL-CM6A-LR and EC-Earth3 models \cite{Jiang2021,Meccia2023}. This focus is motivated not only by previous analysis of these two models, but also by coherence analysis of the AMOC strength by latitude (Supplementary Fig. S3). In all models, AMOC strength in the South Atlantic and equatorial Atlantic lags the AMOC at \ang{40}N, and in all models except one (ACCESS-ESM1-5), the AMOC at 40--\ang{50}N leads that at \ang{40}N. Hence, the northern high latitudes are a plausible driver of centennial-scale AMOC variability in almost all models.

\begin{figure}[htbp]
    \centering
    \includegraphics[width=\textwidth]{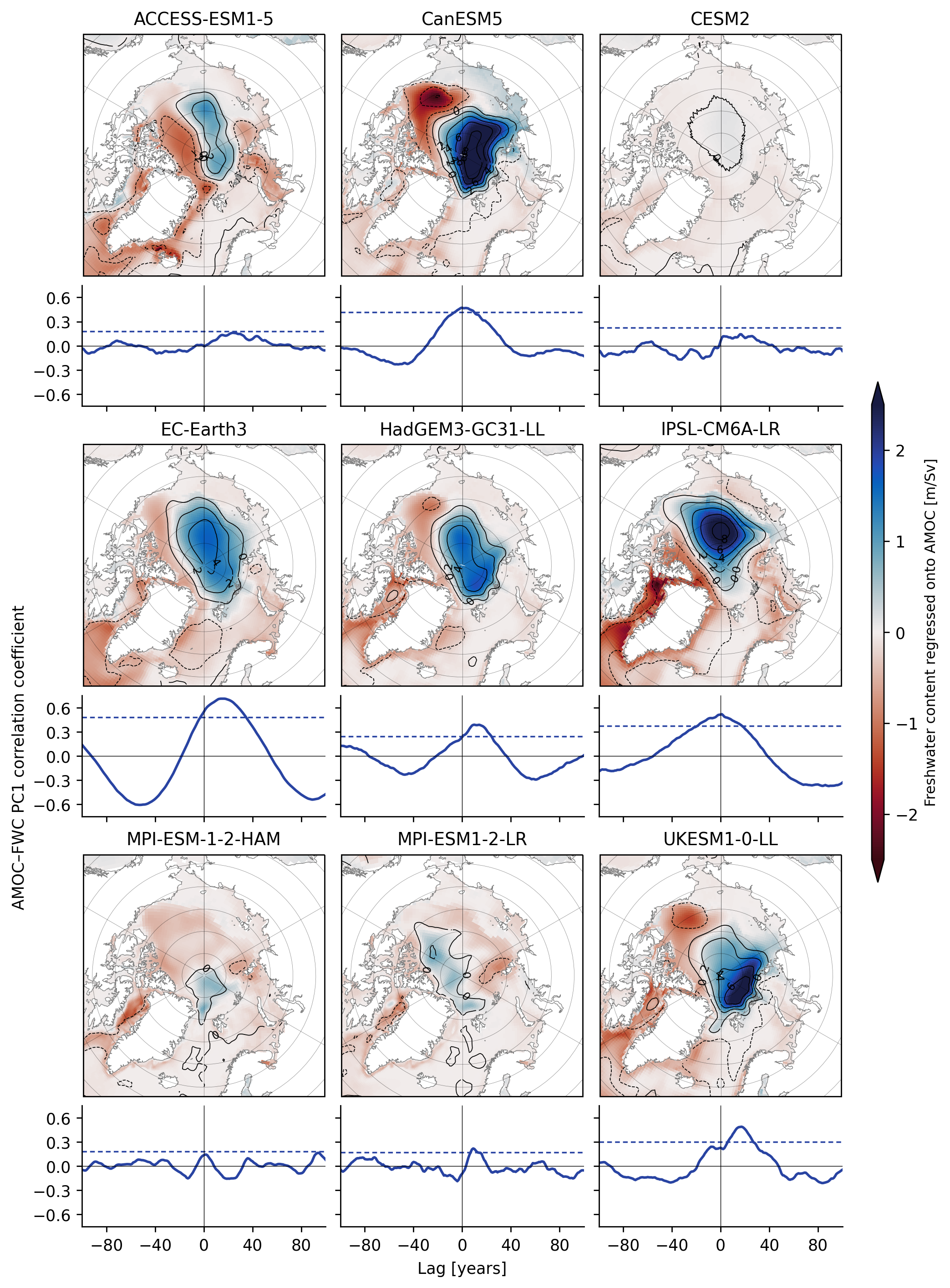}
    \caption{\textbf{Fingerprint of the Arctic--North Atlantic freshwater mechanism.} Maps: Low-pass filtered regression of freshwater content $h_\text{fw}$ (colors, in m/Sv) and sea surface height (contours, in cm/Sv) onto AMOC strength for each model. Line plots: Lagged correlation between AMOC strength and the first principal component of annual mean Arctic Ocean freshwater content. Dashed lines indicate the (one-sided) 95\% confidence level (see Methods). While the maps are based on 70-year low-pass filtered time series, line plots are calculated from unfiltered annual means.}
    \label{fig:freshwater-correlation}
\end{figure}

Fig. \ref{fig:freshwater-correlation} shows one fingerprint of the Arctic--North Atlantic freshwater exchange mechanism -- freshwater content changes in the Arctic Ocean instantaneously regressed onto the 70-year low-pass filtered AMOC. A very similar pattern was shown by \citeA{Jiang2021} and \citeA{Meccia2023} to induce a circulation anomaly that would trap freshwater in the central Arctic Ocean for some decades before releasing it to the North Atlantic and weakening the AMOC. We note that a similar Arctic salinity signature is also shown by \citeA{Jungclaus2005} (their Fig. 9), who proposed a very similar freshwater exchange mechanism except for shorter timescales.

Five of the nine models (IPSL-CM6A-LR, EC-Earth3, HadGEM3-GC31-LL, UKESM1-0-LL, CanESM5) have positive freshwater anomalies exceeding $1\,\text{m}\,\text{Sv}^{-1}$ in the central Arctic Ocean and weaker negative freshwater anomalies elsewhere, in agreement with the pattern in \citeA{Jiang2021} and \citeA{Meccia2023}. These five models match the subset of the ensemble that uses NEMO as its ocean component (``NEMO models'' in the following for simplicity). Among NEMO models, there are differences in the response of the Beaufort Gyre, where HadGEM3-GC31-LL, UKESM1-0-LL and CanESM5 show a negative freshwater anomaly that opposes the central Arctic freshening, while the fresh anomalies in EC-Earth3 and IPSL-CM6A-LR extend towards the Beaufort Gyre. However, the central Arctic anomalies are the dominant contribution to the basin-integrated freshwater content anomaly in all five models (not shown). Among the four models that do not use NEMO, only ACCESS-ESM1-5 shows a similar freshwater regression pattern to the NEMO models when allowing for a 30-year lag behind the AMOC (Supplementary Fig. S5). All other models show only a weak ($< 1\,\text{m}\,\text{Sv}^{-1}$ everywhere) and spatially inhomogeneous Arctic freshwater response to AMOC changes.

The model grouping is supported by empirical orthogonal function (EOF) analysis of the annual mean Arctic freshwater content fields. The first EOF (Supplementary Fig. S6), which explains between 20\% and 50\% of the variance depending on the model, is in good agreement with the regression patterns in Fig. \ref{fig:freshwater-correlation}. The line plots in Fig. \ref{fig:freshwater-correlation} show the amplitude of this first EOF (i.e., the first principal component, PC1) correlated against the AMOC strength at \ang{40}N. In all NEMO models, the maximum of the freshwater content PC1 is in phase or lags the AMOC by up to 20 years, while the lag for ACCESS-ESM1-5 is about 30 years. Strikingly, the correlation between AMOC and the Arctic freshwater content PC1 is significant at the 95\% confidence level for all NEMO models, and the centennial timescale is clearly visible even without applying a low-pass filter. In contrast, in the non-NEMO models, the PC1--AMOC correlation is consistently weaker than in the NEMO models and mostly not significant, and the lagged correlation does not show a clear centennial timescale. Therefore, it is possible that the significant AMOC variability in CESM2 is generated by a mechanism linked to changes at lower latitudes, similar to mechanisms of multi-centennial variability found in its predecessor CESM1 \cite{Li2022,Yang2024a}.

In all models, freshwater anomalies in the Arctic Ocean induce a corresponding change in sea surface height (contours in Fig. \ref{fig:freshwater-correlation}), in line with the expectation that Arctic density anomalies, and therefore steric sea level anomalies, are dominated by salinity changes. While we cannot show velocity vectors for all models because angle information is not available in CMIP6 output, we expect that these sea surface height anomalies induce an anticyclonic geostrophic circulation anomaly as in \citeA{Jiang2021}. This anomaly provides a positive feedback that can prolong the period of the oscillations compared to theoretical expectations that Arctic--North Atlantic inter-basin exchanges should provide oscillations with a multi-decadal period \cite{Wei2022}.

To verify that the freshwater anomalies are indeed a plausible driver of AMOC variability, we evaluate the freshwater transport across Fram Strait. In all NEMO models, the poleward freshwater transport consistently leads the AMOC strength by 20--30 years (Supplementary Fig. S7). Since the mean freshwater transport across Fram Strait is negative (i.e., southward) and dominated by the fresh near-surface East Greenland Current in all models, this implies that the liquid freshwater export from the Arctic Ocean is at its minimum 20--30 years before the AMOC maximum, consistent with the mechanism that an increased southward freshwater transport can weaken the AMOC with a lag, and vice versa \cite<e.g.,>{Dodd2009,Zhang2021b,Wei2022}. In three other models (ACCESS-ESM1-5, CESM2, MPI-ESM1-2-LR) the freshwater transport through Fram strait is also significantly correlated with the AMOC but with a shorter lag (5--10 years). The normalized magnitude (freshwater transport anomaly per Sverdrup of AMOC change) is stronger in the NEMO models and ACCESS-ESM1-5 than in the three remaining models.

\subsection{Sea ice feedbacks amplifying AMOC variability} \label{sec:sea-ice}
\noindent One intriguing similarity across NEMO-based models and ACCESS-ESM1-5 is the normalized (by the magnitude of AMOC variability) magnitude of the freshwater content (Fig. \ref{fig:freshwater-correlation}) and transport (Supplementary Fig. S7), while the absolute magnitude of AMOC variability varies strongly between models (Fig. \ref{fig:amoc-spectra}). This suggests that feedbacks outside of the Arctic Ocean might amplify the centennial-scale AMOC variability in some models.

\begin{figure}[htbp]
    \centering
    \includegraphics[width=\textwidth]{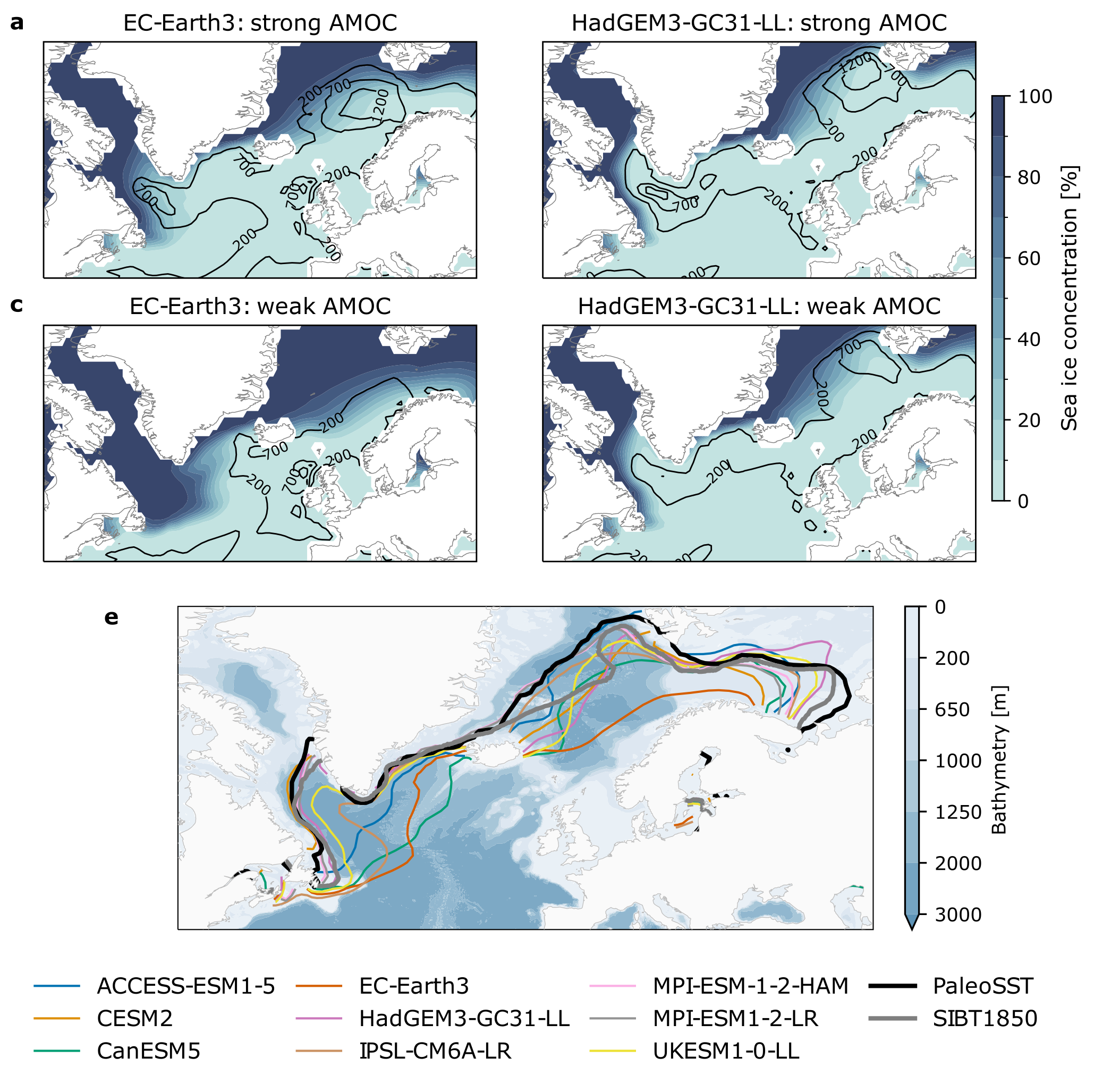}
    \caption{\textbf{Role of the winter sea ice edge for centennial-scale AMOC variability.} (a--d) Composites of mixed layer depth (contours) and sea ice concentration (color shading) in March for two models: (a,c) EC-Earth3 and (b,d) HadGEM3-GC31-LL. Composites are averaged over the strong and weak AMOC phases, which correspond to the intervals in which the low-pass filtered AMOC time series exceeds (Supplementary Fig. S4) plus or minus one standard deviation. Composites for all other models are shown in Supplementary Fig. S8. (e) Mean sea ice edge in March (contours; defined as the contours of 15\% sea ice concentration) for CMIP6 models compared to the PaleoSST reconstruction \cite[black]{Samakinwa2021} and the observational product of \citeA{Walsh2017} (SIBT1850, grey). The bathymetry \cite{GEBCO2023} is shown in the background. Note that the sea ice edge biases in panel (e) are very similar when evaluated over the historical period (1850--2014, Supplementary Fig. S9).}
    \label{fig:sea-ice}
\end{figure}

Here, we show that sea ice cover feedbacks in the Labrador Sea amplify AMOC variability at least in the two models with the strongest centennial-scale variability, EC-Earth3 and IPSL-CM6A-LR. In these models, sea ice in March covers the entire Labrador Sea during a weak AMOC phase, inducing a temporary collapse of Labrador Sea convection \cite[and Fig. \ref{fig:sea-ice}a,c]{Doscher2022}. In contrast, during a strong AMOC phase, the sea ice edge retreats to within the Labrador Sea and the mixed layer reaches more than \SI{700}{m} south of the ice edge. This provides a positive feedback for AMOC strength: weakening of the AMOC cools the North Atlantic, which leads to an extension of sea ice further into the Labrador Sea, which shuts down convective activity near the former ice edge, weakening the AMOC further. Similar feedbacks have been described in the literature for AMOC variability in colder climates \cite<e.g.,>{Klockmann2018}.

Models like HadGEM3-GC31-LL and UKESM1-0-LL, which are characterized by weaker centennial-scale AMOC variability, also show a strong shallowing of the Labrador Sea winter mixed layer during the weak AMOC phase as expected. However, they exhibit little sensitivity of the sea ice edge to the change in AMOC strength (Fig. \ref{fig:sea-ice}b,d and Supplementary Fig. S8).

This difference in feedback strength between models can be linked to biases in the climatological mean position of the winter sea ice edge, defined as the contour of 15\% sea ice concentration in March, in the Labrador Sea (Fig. \ref{fig:sea-ice}e). In HadGEM3, CESM2 and the MPI models, the pre-industrial mean sea ice edge in the Labrador Sea is in proximity to the shelf break. This aligns well with gridded paleoclimate reconstructions for the period 1000--1849 \cite<PaleoSST,>{Samakinwa2021} as well as with the observation-based product of \citeA{Walsh2017} for the period 1850--1880. The proximity to the shelf break means that a retreat further north would not allow for more deep convection due to the shallow bathymetry, while the largest climatological mixed layer depths are located relatively far from the ice edge. In contrast, the mean ice edge position in UKESM1-0-LL, IPSL-CM6A-LR and EC-Earth3 is increasingly biased, reaching far into the Labrador Sea and even into the central North Atlantic in EC-Earth3. This bias allows for the AMOC--mixed layer--sea ice feedback described above, since a sea ice retreat opens areas in which deep convection can form. Two other models (ACCESS-ESM1-5 and CanESM5) also show a strong positive sea ice bias in the Labrador Sea, but do not form deep convection at any time in this region. Instead, their deepest mixed layers in the North Atlantic are east of the Reykjanes Ridge (Supplementary Fig. S8). To summarize, the sea ice bias appears to be a necessary, but not sufficient, condition for strong centennial-scale AMOC variability.

In the Nordic Seas, only EC-Earth3 has a very pronounced positive sea ice bias compared to both reconstructions -- far exceeding that of any other model of the ensemble --, which might explain its largest magnitude of AMOC variability (Fig. \ref{fig:amoc-spectra}). In EC-Earth3, deep convection can shut down simultaneously in the Labrador Sea and the Nordic Seas. Without the presence of deep-water formation in regions sufficiently far from the ice edge (e.g., in the Rockall Trough in EC-Earth3), the sea ice--mixed layer feedback could even lead to a near-shutdown of the AMOC, which has indeed been observed in earlier development versions of EC-Earth3 \cite{Doscher2022} and also IPSL-CM6A-LR \cite{Mignot2021}.

\section{Discussion and Conclusions}
\noindent In this study, we have provided the first systematic multi-model comparison of the magnitude and a mechanism of centennial-scale AMOC variability, using long control simulations from CMIP6 models. Six out of the nine models analyzed exhibit a significant mode of variability at centennial timescales, in line with previous studies which described strong centennial-scale AMOC variability for several individual CMIP6 models \cite{Jiang2021,Waldman2021,Meccia2023}. We showed that a two-way interaction between AMOC strength and Arctic Ocean freshwater content provides a plausible mechanism for centennial-scale AMOC variability at least in the subset of models that use NEMO as their ocean component. Interestingly, it has recently been shown that CMIP6 models that use NEMO simulate stronger Arctic Ocean warming and faster sea ice loss in future projections \cite{Pan2023}. However, whether the strength of North Atlantic and Arctic Ocean variability can formally be linked to stronger sensitivity to future Arctic change -- similar to links between global temperature variability and climate sensitivity \cite{Cox2018,Nijsse2019} -- remains an open question.

While several models in our study show a common mechanism, the magnitude of simulated centennial-scale AMOC variability differs strongly between models, even between those using a similar ocean model configuration. Our results suggest that this diversity is at least partly driven by differences in the sea ice mean state in the Labrador Sea, while recently other (not necessarily independent) mean state biases such as in high-latitude surface density have been shown to contribute as well \cite{Zhao2024}. In our ensemble, only models with a positive winter sea ice bias in the Labrador Sea can produce strong AMOC oscillations, provided that freshwater anomalies are transported in from the Arctic and that the model can temporarily form a deep mixed layer in the Labrador Sea when this sea ice retreats. Why some models instead remain continuously ice-covered and never convect in the Labrador Sea is an open question linked to the more general question of what determines preferential locations of deep convection in climate models \cite{Heuze2021}.

Since the amplitude of unforced centennial-scale AMOC variability is not an observable, selecting models with a realistic sea ice cover could be used to provide an observational constraint. Although such constraints should be corroborated with other lines of evidence, our analysis suggests that the simulated AMOC variability in IPSL-CM6A-LR and EC-Earth3 may be overestimated, which is in line with the results of \citeA{Parsons2020} for global mean surface temperature variability. Alternatively, a more direct comparison of simulated variability with paleoclimate proxies would be possible using simulations of the last millennium \cite{Jungclaus2017}, although volcanic forcing can interfere with the unforced low-frequency variability \cite{ClevelandStout2023}. Unfortunately, very few CMIP6 models have provided last millennium simulations so far.

If the magnitude of centennial-scale AMOC variability indeed depends on the position of the sea ice edge, we would expect a weakening of centennial-scale variability as the sea ice edge retreats northward under global warming, although it might still be significant and even new amplifying mechanisms might become activate \cite{Mehling2023}. Indeed, \citeA{Meccia2023} showed that the amplitude of centennial-scale AMOC variability is strongly reduced in EC-Earth3 when CO$_2$ concentrations are stabilized at different levels above pre-industrial \cite{Fabiano2024}. Whether this is also the case in other models has, to our knowledge, not yet been tested, but our analysis provides a physical mechanism for state-dependence \cite<c.f.>{Bellomo2024}. We also note that similar arguments have been invoked to explain the dependence of millennial-scale variability in the paleoclimate record on background CO$_2$ concentrations \cite[and references therein]{Malmierca-Vallet2023}. State-dependence of centennial-scale AMOC variability would render detection and attribution of AMOC changes more difficult if internal variability is derived from pre-industrial control simulations \cite<e.g.,>{Kelson2022} and is therefore an important topic for future research.

While we used a small but relatively diverse sample of CMIP6 models, one caveat is that all models in this study (and most models in CMIP6) use a relatively coarse ocean resolution of about \ang{1}. Recently, \citeA{Patrizio2023} showed that \ang{1} models are more salinity-stratified in the North Atlantic than their higher-resolution (1/\ang{4}) counterparts. Hence, freshwater anomalies propagating from the Arctic would be expected to influence density anomalies less strongly in higher-resolution models, and the centennial-scale AMOC variability might be weaker in these more realistic setups. In addition, while the transport pathways shown in this study are physically plausible, (lagged) correlations do not demonstrate causation. To this end, future studies could use more physics-based analyses, e.g., through Lagrangian tracers or targeted sensitivity experiments.

To conclude, our results indicate that significant centennial-scale AMOC variability is relatively common among CMIP6 models, but that -- just like on multidecadal timescales \cite{Muir2015,Buckley2016} -- its magnitude varies widely across models. However, process understanding can guide to observables that could aid constraining simulated variability. To this end, our work identified two quantities of interest: the correlation between AMOC strength and Arctic Ocean freshwater content (Section \ref{sec:fw-exchanges}) as well as the mean state of sea ice cover in the North Atlantic (Section \ref{sec:sea-ice}). While the former might still be difficult to observe, sea ice mean-state biases in the North Atlantic could contribute to observationally constrain simulated AMOC variability on timescales beyond the still relatively short observational record.

\section{Open Research}
\noindent All CMIP6 data used in this analysis is freely available from the Earth System Grid Federation (\url{https://esgf-data.dkrz.de/projects/cmip6-dkrz/}). Individual datasets are listed in the Supplementary Information (Table S1). Gridded sea ice data from Walsh et al. were obtained from \url{https://nsidc.org/data/g10010/versions/2} \cite{Walsh2019} and the PaleoSST reconstruction from \url{https://doi.org/10.6084/m9.figshare.c.5369309} \cite{Bronnimann2021}. Code and notebooks to reproduce the diagnostics are available at \url{https://github.com/omehling/centennial-variability-CMIP6} and archived at \url{https://doi.org/10.5281/zenodo.11640570} \cite{MehlingZenodo}.

\acknowledgments
OM and JvH have received funding from the European Union’s Horizon 2020 research and innovation programme under the Marie Sk\l{}odowska-Curie grant agreement No. 956170 (CriticalEarth). KB has received funding from the European Union’s Horizon 2020 research and innovation programme under the Marie Sk\l{}odowska-Curie grant agreement No. 101026907 (CliMOC).

We acknowledge the World Climate Research Programme, which, through its Working Group on Coupled Modelling, coordinated and promoted CMIP6. We thank the climate modeling groups for producing and making available their model output, the Earth System Grid Federation (ESGF) for archiving the data and providing access, and the multiple funding agencies who support CMIP6 and ESGF.

\bibliography{plasim-lsg-amoc,Zenodo_citation}

\clearpage
\includepdf[pages=-,offset=-0.86cm -0.6cm]{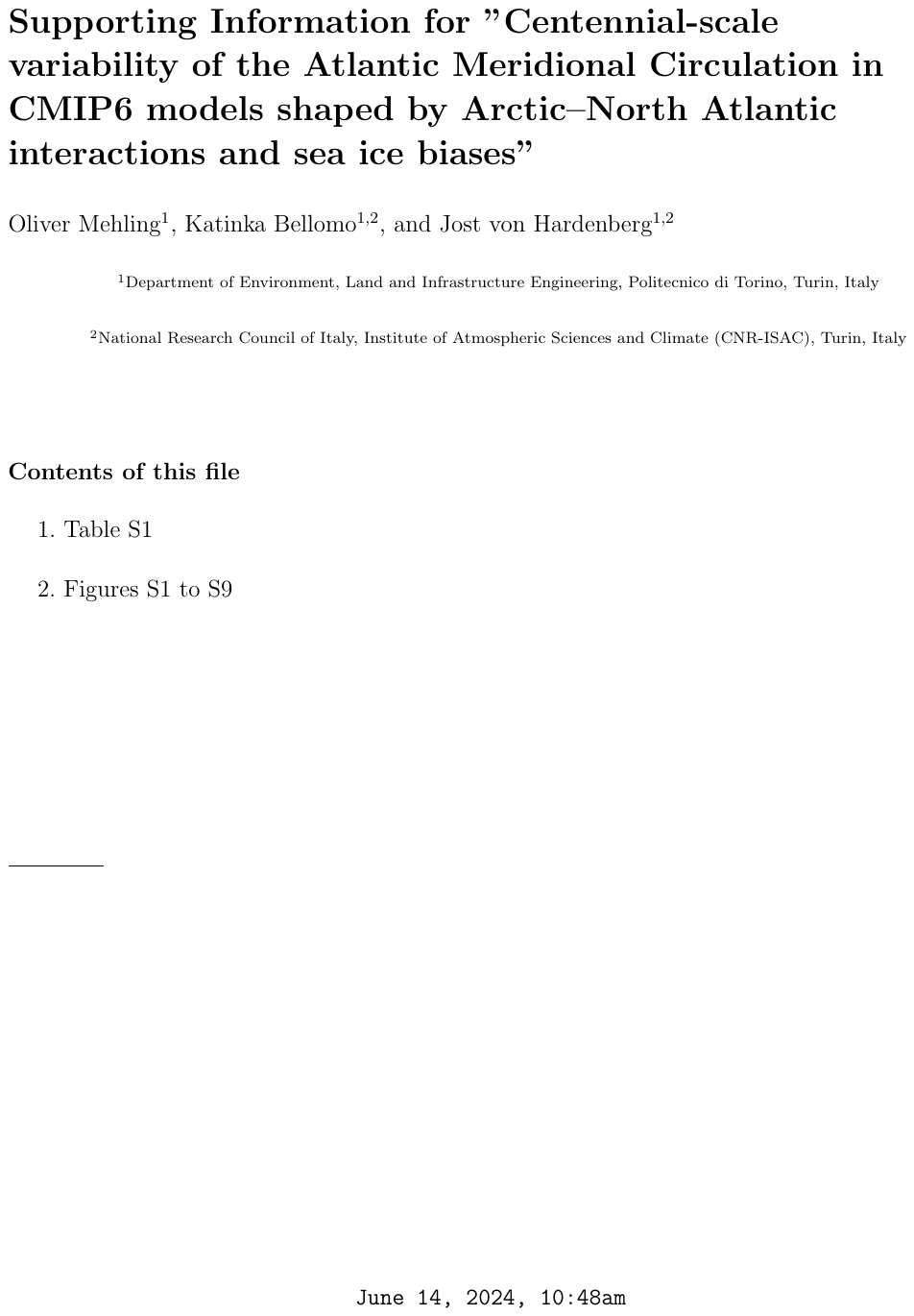}

\end{document}